\documentclass[a4paper,11pt]{article}
\usepackage{pos}
\usepackage{subfig}
\usepackage{caption}
\usepackage{float}
\usepackage{cleveref}
\usepackage{lineno}
\usepackage{setspace}

\title{Monitoring the pointing of the prototype LST-1 using star reconstruction in the Cherenkov camera}
 \ShortTitle{Star tracking} 

\author*[a]{L. Foffano}
\author[a]{A. Carosi}
\author[a]{M. Dalchenko}
\author[a]{M. Heller}
\author[a]{D. Della Volpe}
\author[a]{T. Montaruli}

\affiliation[a]{University of Geneva,  quai Ernest Ansermet 24, Geneva, Switzerland}

\forColl{CTA LST} 

\emailAdd{luca.foffano@unige.ch}
\emailAdd{alessandro.carosi@unige.ch}

\abstract{
The first Large-Sized Telescope (LST-1) proposed for the forthcoming Cherenkov Telescope Array (CTA) has started to operate in 2019 in La Palma. The large structure of LST-1 - with a 23 m mirror dish diameter - imposes a strict control of its deformations that could affect the pointing accuracy and its overall performance. According to CTA specifications that are conceived to resolve e.g. the fine structure of galactic sources, the LST post-calibration pointing accuracy should be better than 14 arcseconds. To fulfill this requirement, the telescope pointing precision is monitored with two dedicated CCD cameras located at the dish center. The analysis of their images allows us to disentangle different systematic deformations of the structure.
In this work, we investigate a complementary approach that offers the possibility to monitor the pointing of the telescope during the acquisition of sky data. After properly cleaning the events from the Cherenkov showers, the reconstructed positions of the stars imaged in the camera field of view are compared to their nominal expected positions in catalogues. This provides a direct measurement of the telescope pointing, that can be used to cross-check the other methods and as a real-time monitoring of the optical properties of the telescope and of the pointing corrections applied by the bending models. Additionally, this method benefits from not relying on specific hardware or dedicated observations.
In this contribution we will illustrate this analysis and show results based on simulations of LST-1.
}
\FullConference{37$^{\rm{th}}$ International Cosmic Ray Conference (ICRC 2021)\\
		July 12th -- 23rd, 2021\\
		Online -- Berlin, Germany}

\begin{document}
\maketitle

\section{Introduction}

\noindent
The {Large-Sized Telescopes} (LSTs) will be the largest telescopes of the Cherenkov Telescope Array (CTA), that represents the next generation of ground-based observatory for the study of very-high-energy (VHE) gamma rays.
The first prototype of LST, called LST-1 \citep{lst-telescope-report-2016, lst-telescope-report-2019}, was inaugurated in October 2018 and is taking commissioning sky data since November 2019.

The LSTs have a parabolic optical reflector of 23~m diameter and a focal length of 28~m \citep{lst-telescope-report-2019}. This provides the LST with a wide reflective surface of about 400~m$^2$ area, thanks to which it will study the Cherenkov showers due to gamma rays with energies as low as  20~GeV. 

Despite their impressive size, the LSTs are built with a light carbon-fiber structure that allows for a fast repositioning of the telescope in order to follow transient events. 
The trade-off between the light structure and the huge size of the telescope implies the unavoidable presence of structure deformations. Such deformations are usually small, but they need to be taken into account in order to achieve the high pointing accuracy that LST has to reach.   

Among the different requirements, the LSTs have to fulfill a pointing accuracy better than 14 arcseconds \citep{lst-telescope-report-2019, lst-tdr}. 
In order to accomplish this, the monitoring of the pointing accuracy is traditionally obtained by disentangling the possible different deformations with specific devices mounted at the dish center  \citep{proceeedings-pointing-2019}:
\begin{itemize}
    \item the starguider camera (SG), a CCD camera that reconstructs every second the pointing direction by comparing the stars in the field of view (FoV) and the center of the camera given by a set of specific LEDs placed around it;
    \item the camera displacement monitor (CDM), a CCD camera measuring at about 10 Hz the displacement of the center of the camera (measured with the LEDs) with respect to the Optical Axis Reference Lasers (OARL);
    \item four distance meters, to verify the precise tilting of the camera. 
\end{itemize}
These devices monitor the deformations and the mispointing during the data taking. Such information is used in the offline data analysis to apply corrections to the data. On the other hand, specific systematic observations to detect the structure deformations overall alt-azimuthal directions are used to apply an online correction to the telescope pointing. Such online corrections rely on the elaboration of a specific {\it bending model} that is applied to the drive system during the observations in order to automatically compensate for deviations due to the structure bending. However, the offline corrections are always needed because such bending model is not intended to correct for other deformations of the telescope structure such as due to changing temperature or wind loads, and these remaining effects are corrected offline with the standard data analysis pipeline.

\section{The star tracking method}

\noindent
All the previously mentioned devices are able to disentangle only partial mis-pointing effects, and their combination is an indirect correction of the existing pointing systematics.

In this work, we apply a complementary and more direct method - called ``star tracking'' method~- to monitor the overall pointing accuracy of the telescope.  Such a method is intended to provide a monitoring of the telescope pointing accuracy and an independent cross-check of the corrections applied with the bending model.
To this end, the star tracking method uses the stars in the FoV during the data taking and contributing to the background of the events. During the observations of a given source, the stars in the field of view (FoV) follow a partial circular trajectory around the average pointing of the telescope. Such trajectories can be used to monitor the average pointing of the telescope (\emph{low-frequency approach}). Alternatively, the information on the stars position can also be used to analyse short time intervals, i.e. when the rotation of the stars is negligible: the comparison between the measured position of the stars in the FoV and their expected positions from the catalogues provides an estimation of the mispointing of the telescope (\emph{high-frequency approach}). \\
The star tracking method offers the following advantages:
\begin{itemize}
    \item It does not require any additional hardware or any specific technical observation time. The method analyses standard data and does not affect the data-taking procedure;
    \item By using Cherenkov events triggered at a frequency of about $8-10$ kHz, the method provides a monitoring of the telescope pointing direction with a frequency comparable and potentially even higher than the standard methods;
    \item Since the information on the stars in the FoV is always contained in the raw data, this method can be applied also retroactively on older data and then provide a historical analysis of the improvement of the telescope pointing precision;
    \item By comparing real data with simulations, the method can also be used for monitoring the optical performance of the telescope, such as the PSF and/or mirror alignment;
    \item Thanks to the high data taking frequency, a future application as an online monitoring tool is being studied.
\end{itemize}
The first applications of such a method were made within the CANGAROO project \citep{kifune_1993, thesis_yoshikoshi_1996, Yoshikoshi:1997}, but more recently also in the CTA framework within the ASTRI project \citep{astri-astrometry} and 
in the pSCT project \citep{psct_paper_2020}. 
In this work we present a systematic application of the method to the PMT camera of the LST-1, that also includes the application of several new implementations and methods (e.g., rotational and translational fit). Additionally, the pipeline made for this work can potentially be adapted to other telescopes, making this software potentially usable as a tool of general utility for different types of telescopes within CTA, even for those not equipped with specific devices for pointing accuracy monitoring. 

The pipeline - that works independently, but is currently being included also in the standard software for the LST data analysis \texttt{cta-lstchain} - has been developed for a systematic analysis of both real data and simulations. {Concerning the resolution of the method, it is correlated to several aspects: the quality of the observations and of the data, but also to how the stars are reconstructed in the pixelized camera, how they are distributed over the camera, their intensity, and other variables. For this reason,} dedicated simulations have been developed in order to test the precision and the robustness of the method under such different cases and under different environmental variables (i.e. the {night-sky background}, the presence of clouds\dots).
Such simulations also aim to confirm that the method satisfies the CTA requirements concerning the pointing accuracy under good-quality dark observations.

\section{The procedure}
\noindent
The method operates on all the events triggered by the telescope. We provide the raw data as an input, and then calibrate them by means of standard camera calibration. For each pixel in the camera, we read the calibrated waveform (still in {analog to digital converter} counts) and compute its variance.
During this process, a cleaning procedure is also applied on a event-by-event basis, balancing the removal of Cherenkov showers without affecting the star light.

\begin{figure}
\centering
\subfloat[Before removing the Cherenkov signal.]{\includegraphics[width=0.35\columnwidth]{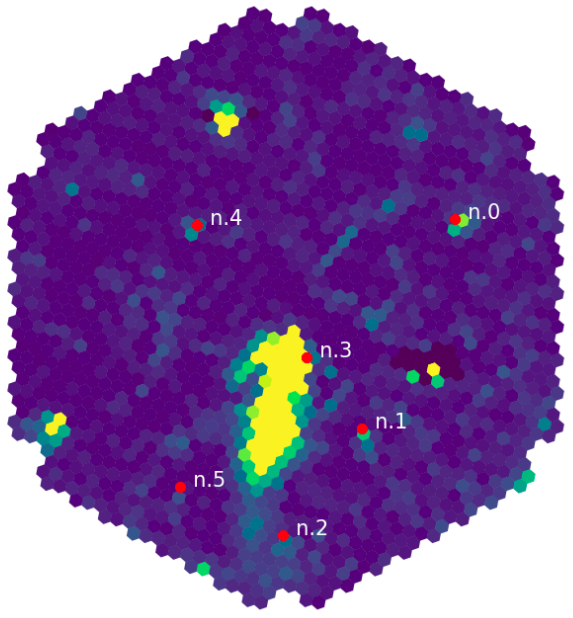}}
\quad
\subfloat[After removing the Cherenkov signal.]{\includegraphics[width=0.35\columnwidth]{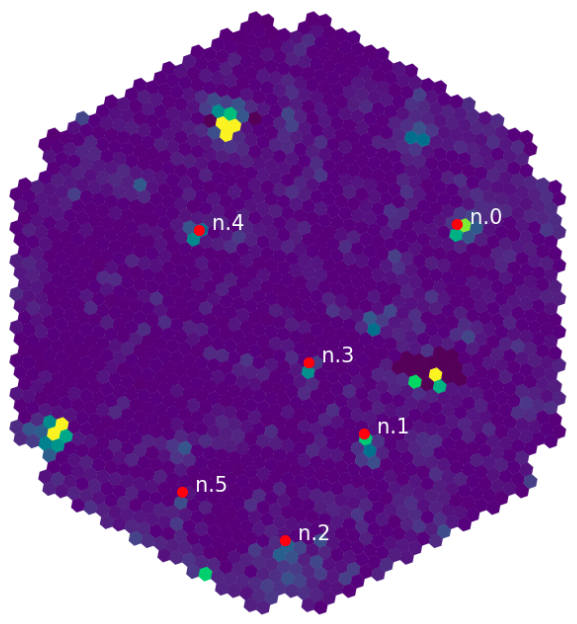}}
\caption{An image coming from real observations before and after the cleaning of the events. The positions of the stars in the FoV (within 2 degrees from the pointing direction) as obtained from the star catalogue at the time of observations are reported. The pixels in the camera are colored depending on the variance of the waveforms in each pixels, as averaged over 200 events: the background is bluish, the excesses are in green to yellow. 
Thanks to the event cleaning and the event stacking to form an image, for example star n.3 - that is strongly affected by some Cherenkov showers - can still be used in the analysis.
Some remaining bright spots after the cleaning may be related to surviving {night-sky background (NSB)} fluctuations or more likely to stars not contained in the chosen catalogue (or not selected).
}
\label{fig:camera_event_and_cleaning}
\end{figure}

In order to increase the signal-to-noise ratio (SNR), the variance of the waveforms of each event is then stacked and averaged over a number of cleaned events $N_{\mathrm{events}}$. This produces an array called \textbf{image} of the camera, to which the program associates time and coordinates information (from the expected pointing direction, given by the drive log file of the telescope).
An example of the improved images after the application of the event-based cleaning algorithm is reported in  \Cref{fig:camera_event_and_cleaning}. 

The expected stars in the FoV are retrieved from the catalog ``The Guide Star Catalog (GSC)`` \citep{Morrison_2001} by providing information about time and pointing direction. 
Given the expected coordinates of the stars $x_{\mathrm{exp}}$, we look for the closest hot-spots in the camera that may correspond to these stars. We also verify the detection of such hot-spots from the background and if the high-voltage of any PMTs of the camera has been reduced due to the star brightness.
Then, we define the background of the camera as the complement of all the pixels associated to the stars, and we extract its average value {\textit{avg\_bkgd}}.

In order to estimate the position of the stars, we compute the center of gravity (CoG) of each star by taking the camera variance of the region hit by the star light and subtracting \textit{avg\_bkgd} on each pixel.
These will be the coordinates $x_{\mathrm{reco}}$ of the reconstructed stars in the camera frame. 
The comparison of the two grid of stars  $x_{\mathrm{exp}}$ and $x_{\mathrm{reco}}$, both with the low-frequency and the high-frequency approach,  provides an estimation of the telescope pointing direction.
The correction of systematic effects due to the star reconstruction (e.g.,  the coma aberration) are taken into account by including them in the coordinates of the expected star positions.


\section{Pointing direction extraction}
\noindent
In our specific application of the method we developed two different approaches, depending mainly on the time scale of the systematics that have to be monitored: the \textit{rotational fit} for the low-frequency approach, and the \textit{translational fit} for the high-frequency approach.

\begin{figure}
\centering
\subfloat[Rotational fit.]{\includegraphics[width=0.35\columnwidth]{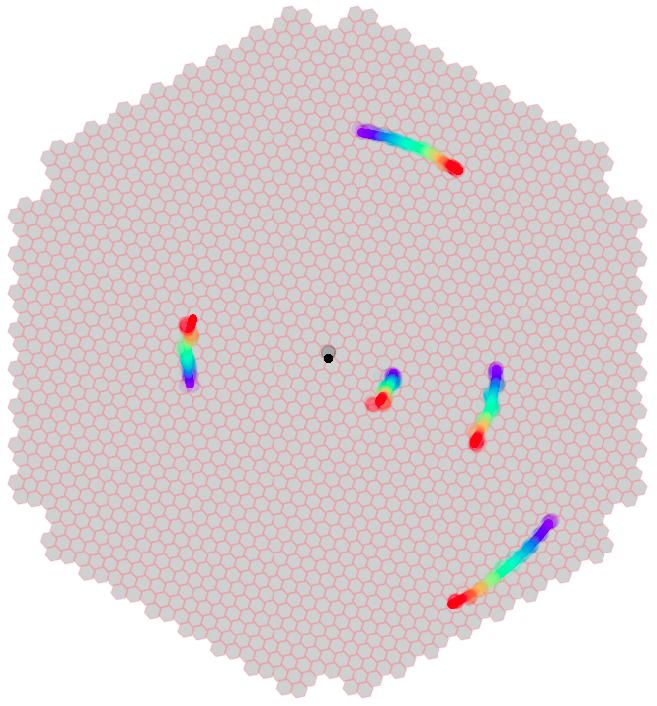}}
\qquad
\subfloat[Translational fit.]{\includegraphics[width=0.51\columnwidth]{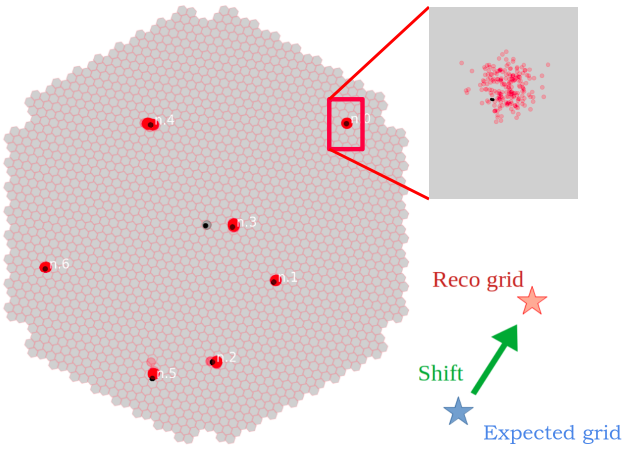}}
\caption{{(a)} A representation of the application of the rotational method for the low-frequency approach. The rotation of the stars inside the FoV is identified and used to estimate the rotation center by means of a rotational fit. {The resulting rotation center of the images (black filled circle)}is assumed to be the average reconstructed telescope position in the camera frame. The color palette from blue to red represents the shift during observations of about 10 min. {(b)}  When the rotation of the stars is negligible, the comparison between the measured position of the stars in the FoV and their expected positions from catalogue is used to provide an estimation of the pointing of the telescope (high-frequency approach). {As represented in the scheme on the right, the translation between the two grids of reconstructed and expected stars is considered.}}
\label{fig:rotational_fit}
\end{figure}

The \textit{rotational fit} is a method conceived to be applied to data sets with a time-scale of several minutes (or more) of observation. During such time, the stars in the FoV are assumed to follow partially a circular trajectory around the average pointing of the telescope, whose radius and length depends on the specific star position in the sky. This implies that  the pointing direction can be estimated by fitting the circular trajectories of the stars and compared to the center of the camera frame (expected pointing direction, given by the drive log file of the telescope).  A representation of this method is shown in \Cref{fig:rotational_fit}.

The \textit{translational fit}  is a method to be applied to data sets with a time-scale of some seconds or less. During this time interval, the stars in the FoV are assumed not to rotate in the FoV, and for this reason the minimization of the distance between the two grids of reconstructed and expected stars is performed with a vectorial term (translation). Current studies show that the frequency of the monitoring may reach at least 4 Hz under good data quality observations.


\section{Simulations}
\noindent
In order to prove the reliability of the method, its accuracy and its precision, dedicated simulations have been performed using the \texttt{sim\_telarray} program \citep{simtelarray}. The importance of such a set of simulations is multi-fold. The reconstructed stars images are indeed the result of the convolution of several factors affecting the star light when passing through the telescope, from optical aberrations to PMT camera efficiency, and such effects have to be taken into account. Furthermore, an optimization of specific analysis parameters is performed to improve the performance and the robustness of the method. 

In order to disentangle all possible systematic effects, the simulations have been subdivided into several types: 
\begin{itemize}
    \item single-star simulations, to reconstruct the systematic effects on the single star reconstruction; 
    \item simplified patterns of stars, to verify the method under simplified configurations of multiple stars and with dedicated scans over several variables (night-sky background, star intensity, number of stars in the FoV\dots);
    \item and finally simulation of real runs, where the real data runs analysed with the program are then simulated (initially with the same configuration, but then also with scans over these variables).
\end{itemize}

\subsection{Estimated precision of the method}
\noindent
In this work, we present the first results of the star tracking method as applied to a set of simulations with simplified patterns of stars. 
The following variables have been considered by evaluating the performance:
\begin{itemize}
    \item variable number of stars in the FoV, between 3 and 6;
    \item variable night-sky background (NSB) level {- i.e. the rate of photo-electrons per pixel due to the NSB -} from 0.01 GHz (virtually no NSB) to 1.5 GHz;
    \item variable offset of the stars from the center of the camera;
    \item variable intensity of the stars.
\end{itemize}
The analysis of the simulations has been performed both with the translational and the rotational fitting method. Additionally, all such configurations have been simulated both with fixed stars in the FoV (for statistical error evaluation) and with moving stars in the FoV with angular velocity similar to the Earth's rotation (as monitoring of the pointing direction over time).

\begin{figure}
\centering
\subfloat[Precision vs star intensity.]{\includegraphics[width=0.7\columnwidth]{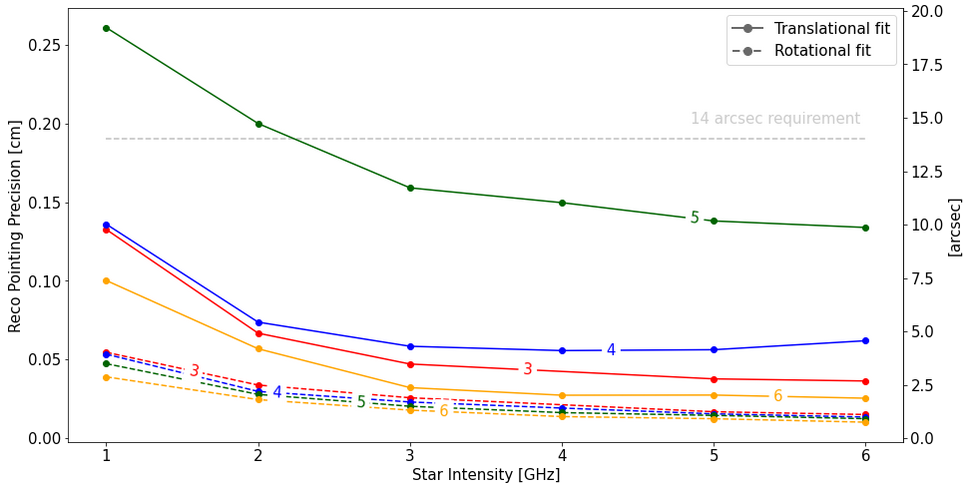}}\\
\subfloat[Precision vs NSB rate. {As a reference, the typical NSB rate for dark night data is between about 0.3 and 0.5 GHz.}]{\includegraphics[width=0.7\columnwidth]{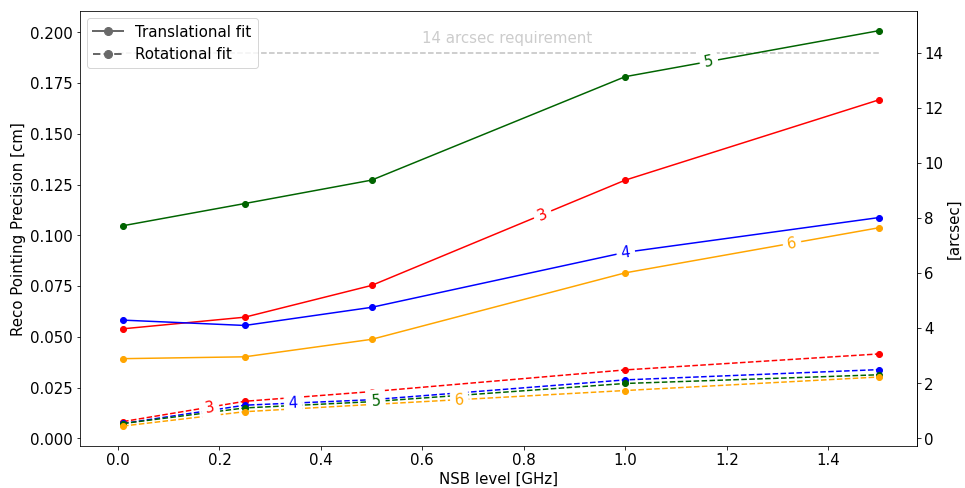}}
\caption{The precision of the method for the configurations of 3, 4, 5 and 6 stars simulated at $1.5^{\circ}$ from the camera center for both the high-frequency (translational fit) and the low-frequency approaches (rotational fit).}
\label{fig:precisionmethod}
\end{figure}

The results of the precision of the method over different stars patterns simulated with a wide range of NSB levels and star intensities are reported in~\Cref{fig:precisionmethod}. {The rotational fit allows us to reach an overall better precision ($\sim 2 ''$) than the translational fit, that shows both a larger average value and a significantly larger spread in the reconstruction precision.} Although not comparable with the resolution of standard pointing monitoring devices (such as the CCD SG camera or the CDM), the achieved resolution for these simplified configurations fulfills the requirement limits requested for CTA/LST pointing precision confirming the method as a valid and complementary alternative for telescopes (not only LST) pointing monitoring. 

Currently, we are investigating further the systematics of this method by simulating more general configuration of stars and relative star intensities.

\section{Conclusions}
\noindent
In this work, we present the systematic application of the star tracking method to real data and simulations of the LST-1 telescope. Such method uses the stars in the field of view during the standard data taking as a monitoring source. Among the several applications, in this work we studied the monitoring of the telescope pointing accuracy. By means of dedicated simulations, we have shown that the precision of the method under good-quality dark observations respect the CTA requirements on the pointing accuracy of the telescope.
Further studies are planned in order to understand the resolution of the method also under more general patterns of stars, by using simulations of real data.

An interesting feature of this method is that, since the information coming from the star light lies in the raw data of the telescope, also a retroactive analysis of the data provides information about the historical improvements of the telescope pointing. 
Thanks to the fact that the method does not require any additional hardware or specific technical observation, the application to other telescopes of the Cherenkov Telescope Array represents an interesting opportunity to be investigated.\\

\noindent
\textbf{Acknowledgements} We gratefully acknowledge financial support from the agencies and organizations listed in \url{www.cta-observatory.org/consortium_acknowledgments}.

\begingroup
\setstretch{0.8}

\bibliographystyle{ieeetr}  
\bibliography{bibliography} 
\endgroup
\setstretch{0.8}
\section*{Full Authors List: \Coll\ Collaboration}
\scriptsize
\noindent
H. Abe$^{1}$,
A. Aguasca$^{2}$,
I. Agudo$^{3}$,
L. A. Antonelli$^{4}$,
C. Aramo$^{5}$,
T.  Armstrong$^{6}$,
M.  Artero$^{7}$,
K. Asano$^{1}$,
H. Ashkar$^{8}$,
P. Aubert$^{9}$
% ...
\\

\noindent
$^{1}$Institute for Cosmic Ray Research, University of Tokyo.
$^{2}$Departament de Física Quàntica i Astrofísica, Institut de Ciències del Cosmos, Universitat de Barcelona, IEEC-UB.
$^{3}$Instituto de Astrofísica de Andalucía-CSIC.
$^{4}$INAF - Osservatorio Astronomico di Roma.
$^{5}$INFN Sezione di Napoli.
$^{6}$Aix Marseille Univ, CNRS/IN2P3, CPPM.
$^{7}$Institut de Fisica d'Altes Energies (IFAE), The Barcelona Institute of Science and Technology.
$^{8}$IRFU, CEA, Université Paris-Saclay.
$^{9}$LAPP, Univ. Grenoble Alpes, Univ. Savoie Mont Blanc, CNRS-IN2P3, Annecy.
% ...
%
%
%

\end{document}